\documentclass[journal,draftclsnofoot,onecolumn,comsoc, a4paper, 12pt, twoside]{IEEEtran}

\usepackage[utf8]{inputenc} 

\usepackage[T1]{fontenc}

\usepackage{cite}

\ifCLASSINFOpdf
 \usepackage[pdftex]{graphicx}
 \graphicspath{{../pdf/}{../jpeg/}}
 \DeclareGraphicsExtensions{.pdf,.jpeg,.png}
\else
 \usepackage[dvips]{graphicx}
 \graphicspath{{../eps/}}
 \DeclareGraphicsExtensions{.eps}
\fi

\usepackage[cmex10]{amsmath}  
\usepackage{amssymb}
\usepackage{algorithmic}

\usepackage{array}

\usepackage{stfloats}

\usepackage{flafter}

\usepackage{url}

\usepackage{cases}

\usepackage{tikz}
\usetikzlibrary{calc} 
\usepackage{pgfplots}
\usetikzlibrary{patterns}
\usetikzlibrary{positioning}
\usetikzlibrary{arrows.meta} 
\usetikzlibrary{shapes} 
\usetikzlibrary{shapes.geometric} 

\usepackage[colorlinks = true]{hyperref}

\RequirePackage{microtype}

\usepackage{xcolor}
\definecolor{TechnionBlue}{rgb}{0, 0.1765, 0.3843} 
\definecolor{OliveGreen}{rgb}{0.4, 0.62, 0.25} 

\usepackage{ifthen}

\usepackage[normalem]{ulem}

\newcommand{\rb}[1] {\left(#1\right)}
\newcommand{\cb}[1] {\left\lbrace #1 \right\rbrace}
\newcommand{\sqb}[1] {\left[#1\right]}

\newcommand*{\nspace}[1] {
	\ifthenelse{\equal{#1}{!}}{\!}{} 
	\ifthenelse{\equal{#1}{tq}}{\kern-1em}{} 
	\ifthenelse{\equal{#1}{mq}}{\mkern-18mu}{} 
}

\newtheorem{proposition}{Proposition}

\usepackage[noamssymbols]{newtxmath}
\usepackage{eucal}

\usepackage[capitalise]{cleveref}
\Crefname{section}{Sec.}{Secs.}

\newcommand{\diag}{\mathop{\mathrm{diag}}}

\begin{document}

\title{Beyond Max-SNR: Joint Encoding for Reconfigurable Intelligent Surfaces}

\author{Roy~Karasik,
	Osvaldo~Simeone,
	Marco~Di~Renzo,
	and~Shlomo~Shamai~(Shitz)
	\thanks{R. Karasik and S. Shamai are with the Department of Electrical Engineering, Technion, Haifa
		32000, Israel (e-mail:roy@campus.technion.ac.il)}
	\thanks{O. Simeone is with the Centre for Telecommunications Research,
		Department of Informatics, King’s College London, London WC2R 2LS, U.K.
		(e-mail: osvaldo.simeone@kcl.ac.uk).}
	\thanks{M. Di Renzo is with the Laboratoire
		des Signaux et Systèmes, CNRS, CentraleSupélec, Univ Paris
		Sud, Université Paris-Saclay, 91192 Gif-sur-Yvette, France (e-mail:
		marco.direnzo@l2s.centralesupelec.fr).}
	\thanks{This work has been supported by the European Research Council (ERC) under the European Union’s Horizon 2020 Research and Innovation Programme (Grant Agreement Nos. 694630 and 725731).}}

\maketitle

\begin{abstract}
	A communication link aided by a Reconfigurable Intelligent Surface (RIS) is studied, in which the transmitter can control the state of the RIS via a finite-rate control link. Prior work mostly assumed a fixed RIS configuration irrespective of the transmitted information. In contrast, this work derives information-theoretic limits, and demonstrates that the capacity is achieved by a scheme that jointly encodes information in the transmitted signal as well as in the RIS configuration.
	In addition, a novel signaling strategy based on layered encoding is proposed that enables practical successive cancellation-type decoding at the receiver. Numerical experiments demonstrate that the standard max-SNR scheme that fixes the configuration of the RIS as to maximize the Signal-to-Noise Ratio (SNR) at the receiver is strictly suboptimal, and is outperformed by the proposed strategies at all practical SNR levels.
\end{abstract}

\section{Introduction}
A Reconfigurable Intelligent Surface (RIS) acts as an ``anomalous mirror'' that can be configured to reflect impinging radio waves towards arbitrary angles, to apply phase shifts, and to modify polarization \cite{ntontin2019reconfigurable}. Due to these desirable properties, RISs are being considered for future wireless networks as means to shape the wireless propagation channel for signal, interference, security, and scattering engineering \cite{ntontin2019reconfigurable,renzo2019smart,wu2019towards,zhao2019survey,chowdhury20196g,zhang2019wireless}. 

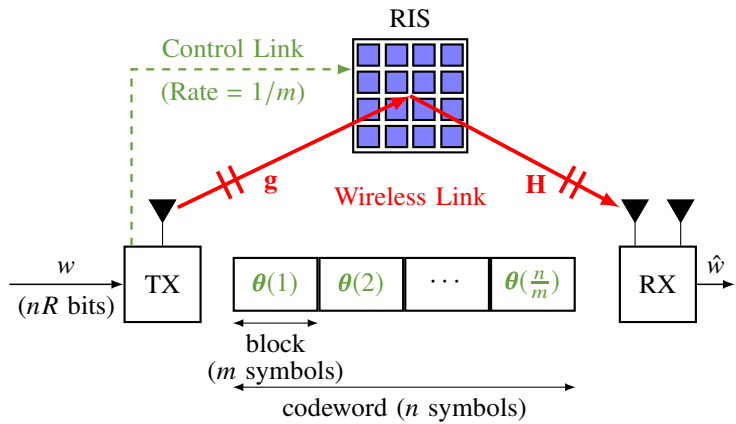
\begin{figure}[!t]
	\centering
	\begin{tikzpicture}[>=latex,
	ant/.style={%
		draw,
		fill,
		regular polygon,
		regular polygon sides=3,
		shape border rotate=180,
	},
]

\node (EN) [thick,draw,minimum width=1cm,minimum height=1cm, font=\small] at (0,0) {TX};
\node (Ant1) [ant, above=0.3cm of EN,scale=0.5] {};
\path[draw] (EN.north) -- (Ant1);

\node (user) [thick,draw,minimum width=1cm,minimum height=1cm,right = 6cm of EN.center, font=\small]  {RX};
\node (Ant2) [ant, above =0.3cm of user,scale=0.5, xshift=-6mm] {};
\path[draw] ($(user.north)-(3mm,0mm)$) -- (Ant2);
\node (Ant3) [ant, above =0.3cm of user,scale=0.5, xshift=6mm] {};
\path[draw] ($(user.north)+(3mm,0mm)$) -- (Ant3);

\node (IRS) [thick,draw,minimum width=1.5cm,minimum height=1.5cm, font=\small] at ($(EN)!0.5!(user)+(0,2.5cm)$) {};
\node (R1) [thick,fill=blue!50,draw,minimum width=0.2cm,minimum height=0.2cm] at ([yshift=5.4mm,xshift=-5.5mm]IRS.center){};
\node (R2) [thick,fill=blue!50,draw,minimum width=0.2cm,minimum height=0.2cm] at ([yshift=-5.4mm,xshift=-5.5mm]IRS.center){};
\node (R3) [thick,fill=blue!50,draw,minimum width=0.2cm,minimum height=0.2cm] at ([yshift=5.4mm,xshift=+5.5mm]IRS.center){};
\node (R4) [thick,fill=blue!50,draw,minimum width=0.2cm,minimum height=0.2cm] at ([yshift=-5.4mm,xshift=+5.5mm]IRS.center){};
\node (R5) [thick,fill=blue!50,draw,minimum width=0.2cm,minimum height=0.2cm] at ([yshift=+5.4mm,xshift=+1.8mm]IRS.center){};
\node (R6) [thick,fill=blue!50,draw,minimum width=0.2cm,minimum height=0.2cm] at ([yshift=+5.4mm,xshift=-1.8mm]IRS.center){};
\node (R7) [thick,fill=blue!50,draw,minimum width=0.2cm,minimum height=0.2cm] at ([yshift=-5.4mm,xshift=+1.8mm]IRS.center){};
\node (R8) [thick,fill=blue!50,draw,minimum width=0.2cm,minimum height=0.2cm] at ([yshift=-5.4mm,xshift=-1.8mm]IRS.center){};
\node (R9) [thick,fill=blue!50,draw,minimum width=0.2cm,minimum height=0.2cm] at ([yshift=1.80mm,xshift=-5.5mm]IRS.center){};
\node (R10) [thick,fill=blue!50,draw,minimum width=0.2cm,minimum height=0.2cm] at ([yshift=-1.8mm,xshift=-5.5mm]IRS.center){};
\node (R11) [thick,fill=blue!50,draw,minimum width=0.2cm,minimum height=0.2cm] at ([yshift=1.8mm,xshift=+5.5mm]IRS.center){};
\node (R12) [thick,fill=blue!50,draw,minimum width=0.2cm,minimum height=0.2cm] at ([yshift=-1.8mm,xshift=+5.5mm]IRS.center){};
\node (R13) [thick,fill=blue!50,draw,minimum width=0.2cm,minimum height=0.2cm] at ([yshift=+1.8mm,xshift=+1.8mm]IRS.center){};
\node (R14) [thick,fill=blue!50,draw,minimum width=0.2cm,minimum height=0.2cm] at ([yshift=+1.8mm,xshift=-1.8mm]IRS.center){};
\node (R15) [thick,fill=blue!50,draw,minimum width=0.2cm,minimum height=0.2cm] at ([yshift=-1.8mm,xshift=+1.8mm]IRS.center){};
\node (R16) [thick,fill=blue!50,draw,minimum width=0.2cm,minimum height=0.2cm] at ([yshift=-1.8mm,xshift=-1.8mm]IRS.center){};
\node [font=\small, above=0.0mm of IRS] {RIS};

\coordinate (A) at ($(Ant1)+(2mm,0)$);
\path[draw,->,line width=0.5mm,red] (A) -- node[below,font=\small,pos=0.4] {$\mathbf{g}$} coordinate[pos=0.2] (P1) coordinate[pos=0.25] (P2) ($(IRS)+(0,0)$);
\draw [red,line width=0.5mm] ($(P1)!0.2cm!270:(A)$) -- ($(P1)!0.2cm!90:(A)$);
\draw [red,line width=0.5mm] ($(P2)!0.2cm!270:(A)$) -- ($(P2)!0.2cm!90:(A)$);
\coordinate (B) at ($(Ant2)+(-2mm,0mm)$);
\path[draw,->,line width=0.5mm,red] ($(IRS)+(0,0)$) -- node[below,font=\small,pos=0.6] {$\mathbf{H}$} coordinate[pos=0.75] (P3) coordinate[pos=0.8] (P4) (B);
\draw [red,line width=0.5mm] ($(P3)!0.2cm!270:(B)$) -- ($(P3)!0.2cm!90:(B)$);
\draw [red,line width=0.5mm] ($(P4)!0.2cm!270:(B)$) -- ($(P4)!0.2cm!90:(B)$);
\node [font=\small, below=3mm of IRS,red] {Wireless Link};

\path[draw,->,line width=0.2mm] ($(EN.west)-(15mm,0)$) -- node[above,font=\small] {$w$} node[below,font=\small] {($nR$ bits)} ($(EN.west)+(0,0)$);
\path[draw,->,line width=0.2mm] ($(user.east)-(0mm,0)$) -- node[above,font=\small] {$\hat{w}$} ($(user.east)+(5mm,0)$);

\draw[line width=0.3mm,OliveGreen,->,dashed] ($(EN.north west)+(3pt,0)$) |- node[above,font=\small,pos=0.73] {Control Link} node[below,font=\small,pos=0.73] {($\text{Rate}=1/m$)} ($(IRS.west)+(0,10pt)$);

\node (t1) [thick,draw,minimum width=1.1cm,minimum height=0.7cm,font=\small] at ($(EN.west)+(2cm,0)$) {\textcolor{OliveGreen}{ $\pmb{\theta}(1)$}};
\node (t2) [right = 0cm of t1,thick,draw,minimum width=1.1cm,minimum height=0.7cm,font=\small] {\textcolor{OliveGreen}{$\pmb{\theta}(2)$}};
\node (dots) [right = 0cm of t2,thick,draw,minimum width=1.1cm,minimum height=0.7cm,font=\small] {$\cdots$};
\node (t3) [right = 0cm of dots,thick,draw,minimum width=1.1cm,minimum height=0.7cm,font=\small] {\textcolor{OliveGreen}{$\pmb{\theta}(\frac{n}{m})$}};
\path [draw,<->] ($(t1.south west)-(0,0.15cm)$) -- node[midway,below]{{\small block}} ($(t1.south east)-(0,0.15cm)$);
\path [] ($(t1.south west)-(0,0.5cm)$) -- node[midway,below]{{\small ($m$ symbols)}} ($(t1.south east)-(0,0.5cm)$);
\path [draw,<->] ($(t1.south west)-(0,1cm)$) -- node[midway,below]{{\small codeword ($n$ symbols)}} ($(t3.south east)-(0,1cm)$);

\end{tikzpicture}
	\caption{Illustration of the network under study consisting of a single-antenna transmitter (TX), a receiver (RX) with $N$ antennas, and a Reconfigurable Intelligent Surface (RIS) with $K$ elements (in the figure, $N=2$ and $K=16$). The transmitter jointly encodes a message $w$ into a codeword of $n$ symbols, sent on the wireless link, and a control action (once every $m$ symbols), sent on the control link to the RIS.}
	\label{fig:simple-model}
\end{figure}

Most prior work, to be reviewed below, proposed to use the RIS as a passive beamformer in order to improve the Signal-to-Noise Ratio (SNR) at the receivers. In contrast, this paper takes a more fundamental information-theoretic perspective on the design of RIS-aided communication links in which the transmitter can control the state of the RIS via a finite-rate control link (see \cref{fig:simple-model}). 
The analysis points to a novel approach of signal engineering via RISs that goes beyond the maximization of the SNR through an information-driven control of the RIS: Rather than being fixed to enhance the SNR, the RIS configuration is jointly encoded with the transmitted signals as a function of the information message.

\textbf{Related Work:} Based on the electromagnetic and physical properties of RISs, free-space path loss models for RIS-aided systems were developed in \cite{ozdogan2019intelligent} and \cite{tang2019wireless}.
The optimization of a fixed RIS configuration has been studied in various scenarios \cite{wu2018Intelligent,zhang2019capacity,perovic2019channel,ntontin2019multi,bjornson2019intelligent,bjornson2019demystifying,guo2019weighted,huang2018energy,han2019intelligent,zhou2019intelligent,huang2018achievable,li2019reconfigurable,abeywickrama2019intelligent,jamali2019intelligent}. We mention here some representative examples. Algorithms for jointly optimizing precoding at the transmitter and beamforming at the RIS were proposed for a point-to-point Multiple-Input Single-Output (MISO) systems in \cite{wu2018Intelligent}, and for Multiple-Input Multiple-Output (MIMO) systems in \cite{zhang2019capacity,perovic2019channel}. RIS-based passive beamforming was compared to conventional relaying methods such as amplify-and-forward and decode-and-forward in \cite{ntontin2019multi,bjornson2019intelligent}, and to multi-antenna systems in \cite{bjornson2019demystifying}. 
Algorithms for maximizing weighted sum-rate and energy efficiency in an RIS-aided multi-user MISO systems were proposed in \cite{guo2019weighted} and \cite{huang2018energy,han2019intelligent}, respectively. A multi-group multi-cast RIS-aided system was studied in \cite{zhou2019intelligent}, and efficient algorithms were proposed to maximize the sum-rate achieved by all groups. 

To the best of our knowledge, the only paper that considers joint encoding of the transmitted signal and RIS state is \cite{basar2019large}, in which the receiver antenna for which SNR is maximized encodes the information bits using index modulation \cite{khandani2013media}. However, reference \cite{basar2019large} does not address the optimality of the proposed index modulation scheme. Moreover, the proposed scheme fixes the configuration of the RIS for the entire duration of the transmission, and hence it provides minor rate increments for large coding blocks.

\textbf{Main Contributions:} In this paper, we study the RIS-aided system with a single-antenna transmitter and a receiver with $N$ antennas illustrated in \cref{fig:simple-model}. We first derive the capacity of the system for any RIS control rate, and prove that joint encoding of transmitted signals and RIS configuration is generally necessary to achieve the maximum information rate. In addition, we explicitly characterize the performance gain of joint encoding in the high-SNR regime.
Then, we propose an achievable scheme based on layered encoding and Successive Cancellation Decoding (SCD) that enable information encoding in the RIS configuration, while supporting standard separate encoding and decoding strategies. Numerical experiments demonstrate that, for SNR levels of practical interest, capacity-achieving joint encoding provides significant gain over the max-SNR approach.

\textbf{Notation:} For any positive integer $K$, we define the set $[K]\triangleq \{1,2,\ldots,K\}$. For any real number $\gamma$, we define $(\gamma)^+\triangleq\max\{0,\gamma\}$. The cardinality of a set $\mathcal A$ is denoted as $|\mathcal{A}|$. The $\ell^2$-norm of a vector $\mathbf{z}$ is denoted as $\lVert\mathbf{z}\rVert$ and the Frobenius norm of a matrix $\mathbf{Z}$ is denoted as $\lVert\mathbf{Z}\rVert_\text{F}$. $\diag(\mathbf{x})$ represents a diagonal matrix with diagonal given by vector $\mathbf{x}$.

\section{System Model}\label{sec:model}
We consider the system depicted in \cref{fig:simple-model} in which a Reconfigurable Intelligent Surface (RIS) of $K$ elements is leveraged by a single-antenna transmitter for the purpose of enhancing its communication with a receiver equipped with $N$ antennas over a quasi-static fading channel. 
A coding block consists of $n$ symbols taken from a constellation $\mathcal{B}$ of $B=|\mathcal B|$ points, and is used to communicate a message $w\in[1,\ldots,2^{nR}]$ of $nR$ bits with rate $R$ [bits/symbol].
Unlike most prior work, the transmitter encodes message $w$, not only into the codeword of $n$ symbols sent on the wireless link to the receiver, but also jointly in the configuration of the RIS. The latter is defined by the phase shifts that each of the $K$ RIS elements applies to the impinging wireless signal. 

Following \cite{wu2019towards}, we assume that the phase shift applied by each element is chosen from a finite set $\mathcal A$ of $A=|\mathcal A|>0$ distinct hardware-determined values. Moreover, there is a limit on the rate at which the RIS can be controlled, such that the phase shifts $\pmb{\theta}(t)$ are fixed for \emph{blocks} of $m>0$  consecutive transmitted symbols. That is, as illustrated in \cref{fig:simple-model}, the state $\pmb{\theta}(t)$ of the RIS can be changed only at the beginning of each block $t\in[n/m]$ of $m$ transmitted symbols. Note that if $m=n$, the configuration of the RIS is fixed for the entire duration of the transmission as assumed in the prior work reviewed above. We take $n$ to be a multiple of $m$.

We assume that the direct link between transmitter and receiver is blocked, as in, e.g., \cite{bjornson2019demystifying}, so that propagation from transmitter to receiver occurs through reflection from the RIS. Let the received signal matrix $\mathbf{Y}(t)=(\mathbf{y}_1(t),\ldots,\mathbf{y}_m(t))\in\mathbb C^{N\times m}$ collect the received samples in the $t$th block of a codeword, so that column $\mathbf{y}_i(t)$, $i\in[m]$, denotes the signal received at the $N$ antennas for the $i$th transmitted symbol in the block. The signal received on the wireless link within each $t$th block, with $t\in[n/m]$, can then be written as
\begin{IEEEeqnarray}{c}\label{eq:channel}
	\mathbf{Y}(t)=\mathbf{H}\mathbf{S}(t)\mathbf{g}\mathbf{x}(t)+\mathbf{Z}(t),
\end{IEEEeqnarray}
where the transmitted signal $\mathbf{x}(t)=(x_1(t),\ldots,x_{m}(t))\in\mathcal B^{1\times m}$ consists of the $m$ symbols transmitted in the $t$th block; 
the channel vector $\mathbf{g}\in\mathbb C^{K\times 1}$ denotes the quasi-static flat-fading channel from the transmitter to the RIS;
the RIS configuration matrix 
\begin{IEEEeqnarray}{c}
	\mathbf{S}(t)=\diag\left(e^{j\theta_1(t)},\ldots,e^{j\theta_K(t)}\right)
\end{IEEEeqnarray}
denotes the phase shifts applied by the RIS during the transmission of the $t$th block, with $\theta_k(t)\in\mathcal A$ denoting the phase-shift for the $k$th RIS element, $k\in[K]$; 
the channel matrix $\mathbf{H}\in\mathbb C^{N\times K}$ denotes the quasi-static flat-fading channel from the RIS to the $N$ receiver antennas; and the white Gaussian noise matrix $\mathbf{Z}(t)\in\mathbb C^{N\times m}$, whose elements are independent and identically distributed (i.i.d.) as $\mathcal{CN}(0,1)$, denotes the additive noise at the receiving antennas during the transmission of the $t$th block.
The transmitted signal is subject to the power constraint $\mathbb E[|x_i(t)|^2]\leq P$ for $i\in[m]$ and some $P>0$.
We assume full Channel State Information (CSI) in the sense that the transmitter and receiver know the quasi-static channel vectors $\mathbf{g}$ and $\mathbf{H}$, which remain fixed throughout the $n$ symbols corresponding to the transmission of a message $w$.

Based on the message $w$ and CSI given by the pair $(\mathbf{g},\mathbf{H})$, the encoder jointly selects a codeword $\mathbf{x}(t)$, as well as a sequence $\pmb{\theta}(t)=(\theta_1(t),\ldots,\theta_K(t))$ of RIS configurations, for $t\in[n/m]$.
Having received signal $\mathbf{Y}(t)$ in~\eqref{eq:channel} for $t\in[n/m]$, the decoder produces the estimate $\hat{w}=\hat{w}(\mathbf{Y}(1),\ldots, \mathbf{Y}(n/m),\mathbf{g},\mathbf{H})$ using knowledge of the CSI. As in the conventional definition in information theory (see, e.g., \cite[Ch. 7]{cover2006elements}), a rate $R(\mathbf{g},\mathbf{H})$ is said to be \emph{achievable} if the probability of error satisfies the limit $\Pr(\hat{w}\neq w)\rightarrow 0$ when the codeword length grows large, i.e., $n\rightarrow\infty$. The corresponding capacity $C(\mathbf{g},\mathbf{H})$ is defined as the maximum over all achievable rates, i.e.,
\begin{IEEEeqnarray}{c}
	C(\mathbf{g},\mathbf{H})\triangleq\sup\{R(\mathbf{g},\mathbf{H}):R(\mathbf{g},\mathbf{H})\text{ is achievable}\},
\end{IEEEeqnarray}
where the supremum is taken over all joint encoding and decoding schemes.
Finally, we define the \emph{average rate and capacity} as 
\begin{IEEEeqnarray}{c}
	R\triangleq\mathbb E[R(\mathbf{g},\mathbf{H})]\text{ and }C\triangleq\mathbb E[C(\mathbf{g},\mathbf{H})],
\end{IEEEeqnarray}
where the average is taken over the distribution of the CSI $(\mathbf{g},\mathbf{H})$.

\section{Joint Encoding: Channel Capacity}
In this section, we derive the capacity $C(\mathbf{g},\mathbf{H})$ and we argue that the standard max-SNR method that does not encode information in the RIS configuration (see, e.g., \cite{wu2018Intelligent,zhang2019capacity,perovic2019channel,ntontin2019multi,bjornson2019intelligent,bjornson2019demystifying,guo2019weighted,huang2018energy,han2019intelligent,zhou2019intelligent,huang2018achievable,li2019reconfigurable,abeywickrama2019intelligent,jamali2019intelligent}) is strictly suboptimal.
\begin{proposition}\label{prop:capacity}
	The capacity of the channel~\eqref{eq:channel} is given as
	\begin{IEEEeqnarray}{c}\label{eq:capacity}
		C(\mathbf{g},\mathbf{H}) = -N\log_2(e)
		-\frac{1}{m}\min_{\substack{
				p(\mathbf{x},\pmb{\theta}):\\
				\mathbb E[|x_i|^2]\leq P,\\
				\mathbf{x}\in\mathcal B^m,~\pmb{\theta}\in\mathcal A^K
		}}
		\sum_{\mathbf{x}\in\mathcal B^m}
		\sum_{\pmb{\theta}\in\mathcal A^K}p(\mathbf{x},\pmb{\theta})\mathbb E \sqb{f_\text{c}(\mathbf{x},\pmb{\theta},\mathbf{Z})},
	\end{IEEEeqnarray}
	where we have defined function
	\begin{IEEEeqnarray}{c}\label{eq:def_f_joint}
		f_\text{c}(\mathbf{x},\pmb{\theta},\mathbf{Z})\triangleq 
		 \log_2\rb{\sum_{\mathbf{x}'\in\mathcal B^m}\sum_{\pmb{\theta}'\in\mathcal A^K}p(\mathbf{x}',\pmb{\theta}')\exp\left(-\left\lVert \mathbf{Z}+\mathbf{H}(\mathbf{S}\mathbf{g}\mathbf{x}-\mathbf{S}'\mathbf{g}\mathbf{x}')\right\rVert^2_\text{F}\right)}
	\end{IEEEeqnarray}
	with matrices $\mathbf{S}=\diag(\exp(j\theta_1),\ldots,\exp(j\theta_K))$ and $\mathbf{S}'=\diag(\exp(j\theta'_1),\ldots,\exp(j\theta'_K))$, and the expectation in~\eqref{eq:capacity} being taken with respect to a matrix $\mathbf{Z}$ whose elements are i.i.d. as $\mathcal{CN}(0,1)$.
\end{proposition} 
\begin{IEEEproof}
	See Appendix \ref{app:capacity}.
\end{IEEEproof}

At a computational level, problem~\eqref{eq:capacity} is convex (see Appendix \ref{app:capacity}), and hence it can be solved using standard tools. 
In terms of the operational significance of \cref{prop:capacity}, achieving capacity~\eqref{eq:capacity} generally requires joint encoding over codeword symbols $\mathbf{x}$ and RIS configuration variables $\pmb{\theta}$, as well as joint decoding of message $w$ based on information encoded over both $\mathbf{x}$ and $\pmb{\theta}$ at the receiver. This is reflected in~\eqref{eq:capacity} in the optimization over the joint distribution $p(\mathbf{x},\pmb{\theta})$. 
In the next section, we will consider a suboptimal approach that uses separate encoding and decoding over $\mathbf{x}$ and $\pmb{\theta}$ through layering.

The following proposition derives achievable rates for the standard max-SNR approach \cite{wu2018Intelligent,zhang2019capacity,perovic2019channel,ntontin2019multi,bjornson2019intelligent,bjornson2019demystifying,guo2019weighted,huang2018energy,han2019intelligent,zhou2019intelligent,huang2018achievable,li2019reconfigurable,abeywickrama2019intelligent,jamali2019intelligent}, whereby the state of the RIS $\pmb{\theta}$ is fixed for the entire transmission, irrespective of message $w$, so as to maximize the SNR at the receiver.
\begin{proposition}\label{prop:max_snr}
	The rate 
	\begin{IEEEeqnarray}{c}\label{eq:maxSNR_discrete}
		R_\text{max-SNR}(\mathbf{g},\mathbf{H})=
		-N\log_2(e)
		-\min_{\substack{
				p(x),~\pmb{\theta}:\\
				\mathbb E[|x|^2]\leq P,\\
				x\in\mathcal B,~\pmb{\theta}\in\mathcal A^K
		}}
		\sum_{x\in\mathcal {B}}p(x)\mathbb E\sqb{f_\text{max-SNR}(x,\pmb{\theta},\mathbf{z})}
	\end{IEEEeqnarray}
	is achievable by selecting the phase shift vector $\pmb{\theta}$ so as to maximize SNR at the receiver, where we have defined function 
	\begin{IEEEeqnarray}{c}\label{eq:def_f(x,z)}
		f_\text{max-SNR}(x,\pmb{\theta},\mathbf{z})\triangleq \log_2\rb{\sum_{x'\in\mathcal B}p(x')\exp\left(-\left\lVert \mathbf{z}+\mathbf{H}\mathbf{S}\mathbf{g}(x-x')\right\rVert^2\right)},
	\end{IEEEeqnarray}
	and the expectation in~\eqref{eq:maxSNR_discrete} is taken with respect to $\mathbf{z}\sim\mathcal{CN}(0,\mathbf{I}_N)$.
\end{proposition}
\begin{IEEEproof}
	See Appendix \ref{app:maxSNR_discrete}.
\end{IEEEproof}

The max-SNR rate~\eqref{eq:maxSNR_discrete} can be again computed using convex optimization tools. It is generally smaller than the counterpart capacity~\eqref{eq:capacity}, and we will evaluate the corresponding performance loss in \cref{sec:numerical} via numerical experiments. The following proposition settles the comparison in the high-SNR regime.
\begin{proposition}\label{prop:high_SNR}
	For any finite input constellation $\mathcal B$, the high-SNR limit of the average capacity is given as
	\begin{IEEEeqnarray}{c}\label{eq:lim_high_snr}
		\lim_{P\rightarrow\infty}C=\frac{\log_2(|\mathcal C|)}{m},
	\end{IEEEeqnarray}
	where we have defined the set
	\begin{IEEEeqnarray}{c}
		\mathcal C\triangleq \cb{\mathbf{C}:\mathbf{C}=\rb{e^{j\theta_1},\ldots,e^{j\theta_K}}^\intercal\mathbf{x},~ \mathbf{x}\in\mathcal B^{m},~\pmb{\theta}\in\mathcal A^K}
	\end{IEEEeqnarray}
	containing all possible distinct configurations of the matrix $\mathbf{C}=(e^{j\theta_1},\ldots,e^{j\theta_K})^\intercal\mathbf{x}$.
	Furthermore, for a given cardinality $B=|\mathcal B|$ of the constellation, the limit~\eqref{eq:lim_high_snr} is maximized for the Amplitude Shift Keying (ASK) constellation
	\begin{IEEEeqnarray}{c}\label{eq:ASK_mod}
		\mathcal B=\{\beta,3\beta,\ldots,(2B-1)\beta\},
	\end{IEEEeqnarray}
	where the factor $\beta\triangleq\sqrt{3P/[3+4(B^2-1)]}$ ensures average power $P$, yielding the limit
	\begin{IEEEeqnarray}{c}\label{eq:lim_high_snr_ASK}
		\lim_{P\rightarrow\infty}C=\log_2(B)+\frac{K\log_2(A)}{m}.
	\end{IEEEeqnarray}
\end{proposition}
\begin{IEEEproof}
	See Appendix \ref{app:high_SNR}.
\end{IEEEproof}

In the high-SNR regime, the rate of the max-SNR scheme is limited to $\log_2(B)$. Therefore,
\cref{prop:high_SNR} demonstrates that, for any RIS configuration set $\mathcal A$ of $A$ distinct phases, in the high-SNR regime, modulating the RIS state can be used to increase the achievable rate by $K\log_2(A)/m$ bits per symbol as compared to the max-SNR scheme. Furthermore, the proposition shows that, in this regime, it is optimal to use the ASK modulation~\eqref{eq:ASK_mod}, and it implies that choosing independent codebooks for input $\mathbf{x}$ and RIS configuration $\pmb{\theta}$, i.e., setting $p(\mathbf{x},\pmb{\theta})=p(\mathbf{x})p(\pmb{\theta})$ in~\eqref{eq:capacity}, does not cause any performance loss at high SNR. As an additional comparison, we note that the high-SNR performance of the index modulation scheme proposed in \cite{basar2019large} is upper-bounded by $\log_2(B)+\log_2(N)/n$. Therefore, its suboptimality as compared to~\eqref{eq:lim_high_snr_ASK} increases with the ratio $n/m$.

\section{Layered Encoding}\label{sec:layered}
As discussed, achieving the capacity~\eqref{eq:capacity} requires jointly encoding the message over the phase shift vector $\pmb{\theta}$ and the transmitted signal $\mathbf{x}$, while performing optimal, i.e., maximum-likelihood joint decoding at the receiver. In this section we propose a strategy based on layered encoding and Successive Cancellation Decoding (SCD) that uses only standard separate encoding and decoding strategies, while still benefiting from the modulation of information over the state of the RIS to improve over the max-SNR approach.

To this end, the message $w$ is split into two sub-messages, or layers, $w_1$ and $w_2$, such that $w_1$, of rate $R_1$, is encoded by the phase shift vector $\pmb{\theta}$, whereas $w_2$, of rate $R_2$, is encoded by the transmitted signal $\mathbf{x}=(x_1,\ldots,x_m)$. In order to enable decoding using standard SCD, the first $\tau$ symbols $x_1,\ldots,x_\tau$, with $\tau\geq 1$, in vector $\mathbf{x}$ are fixed and used as pilots. 
The receiver starts by decoding $w_1$ using the first $\tau$ vectors $\mathbf{y}_1,\ldots,\mathbf{y}_\tau$, in every received block $\mathbf{Y}=(\mathbf{y}_1,\ldots,\mathbf{y}_m)$. This allows the decoder to obtain vector $\pmb{\theta}$, which is then used to decode $w_2$. This strategy achieves the rate detailed in \cref{prop:ach_layered_discrete}.
\begin{proposition}\label{prop:ach_layered_discrete}
	A strategy based on layered encoding and SCD achieves the rate
	\begin{IEEEeqnarray}{c}\label{eq:ach_layered_discrete}
		R_\text{layered}(\mathbf{g},\mathbf{H},\tau)= \frac{1}{\tilde{m}}R_1(\mathbf{g},\mathbf{H},\tau)+\frac{\tilde{m}-\tau}{\tilde{m}}R_2(\mathbf{g},\mathbf{H}),
	\end{IEEEeqnarray}
	where $\tilde{m}\triangleq\max\{\tau+1,m\}$; rate $R_1(\mathbf{g},\mathbf{H},\tau)$ is defined as
	\begin{IEEEeqnarray}{c}\label{eq:def_R1}
		R_1(\mathbf{g},\mathbf{H},\tau)\triangleq K\log_2(A)-N\log_2(e)
		-\sum_{\pmb{\theta}\in \mathcal A^K}\frac{\mathbb E\sqb{f_1(\pmb{\theta},\mathbf{z})}}{A^K}\IEEEeqnarraynumspace
	\end{IEEEeqnarray}
	with function
	\begin{IEEEeqnarray}{c}\label{eq:def_f1}
		f_1(\pmb{\theta},\mathbf{z})\triangleq\log_2\rb{\sum_{\pmb{\theta}'\in\mathcal{A}^K}\exp\left(-\lVert \mathbf{z}+\sqrt{\tau P}\mathbf{H}(\mathbf{S}-\mathbf{S'})\mathbf{g}\rVert^2\right)}\IEEEeqnarraynumspace
	\end{IEEEeqnarray}
	and matrices $\mathbf{S}=\diag(\exp(j\theta_1),\ldots,\exp(j\theta_K))$ and $\mathbf{S}'=\diag(\exp(j\theta'_1),\ldots,\exp(j\theta'_K))$; and rate $R_2(\mathbf{g},\mathbf{H})$ is defined as
	\begin{IEEEeqnarray}{c}\label{eq:tildeR2}
		R_2(\mathbf{g},\mathbf{H})\triangleq\log_2(B)-N\log_2(e)
		-\sum_{x\in\mathcal B}\sum_{\pmb{\theta}\in\mathcal A^K}\frac{\mathbb E\sqb{f_2(x,\pmb{\theta},\mathbf{z})}}{B\cdot A^K}\IEEEeqnarraynumspace
	\end{IEEEeqnarray}
	with function
	\begin{IEEEeqnarray}{l}
		f_2(x,\pmb{\theta},\mathbf{z})\triangleq
		\log_2\rb{\sum_{x'\in\mathcal B}\exp\rb{-\left\lVert \mathbf{z}+\sqrt{\alpha(\pmb{\theta})}\mathbf{H}\mathbf{S}\mathbf{g}(x-x')\right\rVert^2}}\IEEEeqnarraynumspace
	\end{IEEEeqnarray}
	and a power allocation parameter
	\begin{IEEEeqnarray}{c}\label{eq:power_allocation}
		\alpha(\pmb{\theta})\triangleq \rb{\frac{1}{\gamma_0}-\frac{1}{\gamma(\pmb{\theta})}}^+,
	\end{IEEEeqnarray}
	which depends on the SNR
	\begin{IEEEeqnarray}{c}
		\gamma(\pmb{\theta})\triangleq\left\lVert\mathbf{H}\diag\left(e^{j\theta_1},\ldots,e^{j\theta_K}\right)\mathbf{g}\right\rVert^2P
	\end{IEEEeqnarray}
	and on a cutoff parameter $\gamma_0$ satisfying the equality
	\begin{IEEEeqnarray}{c}\label{eq:avg_power_eq}
		\frac{1}{A^K}\sum_{\pmb{\theta}\in\mathcal A^K}\alpha(\pmb{\theta})=\frac{B\cdot P}{\sum_{x\in\mathcal B}|x|^2}.
	\end{IEEEeqnarray}
	The expectations in~\eqref{eq:def_R1} and~\eqref{eq:tildeR2} are taken with respect to random vector $\mathbf{z}\sim\mathcal{CN}(0,\mathbf{I}_N)$.	
\end{proposition} 
\begin{IEEEproof}
	Rate $R_1(\mathbf{g},\mathbf{H},\tau)$ is obtained by modulating the RIS phases, and follows in a manner similar to \cite[Eq. (4)]{he2005Computing}. Rate $R_2(\mathbf{g},\mathbf{H})$ is instead obtained by applying ``water-filling'' power allocation \cite{goldsmith1997Capacity}.
	Details can be found in Appendix \ref{app:ach_layered}.
\end{IEEEproof}

\section{Numerical Results}\label{sec:numerical}
In this section, we provide numerical examples with the main aim of comparing the capacity~\eqref{eq:capacity} with the rate~\eqref{eq:maxSNR_discrete} achieved by the max-SNR approach and the rate~\eqref{eq:ach_layered_discrete} of the layered-encoding strategy. For the set of RIS configurations, we assume $A$ uniformly spaced phases in the set $\mathcal A\triangleq\{0,2\pi/A,\ldots,2\pi(A-1)/A\}$, whereas, for the input constellation, we consider ASK and Phase Shift Keying (PSK). All rates are averaged over the channel vector $\mathbf{g}\sim\mathcal{CN}(0,\mathbf{I}_K)$ and channel matrix $\mathbf{H}$ whose elements are i.i.d. as $\mathcal{CN}(0,1)$ and independent of $\mathbf{g}$.

\begin{figure}[!t]
	\centering
	\resizebox {0.5\columnwidth} {!} {	
%
%
\definecolor{mycolor1}{rgb}{0.00000,0.44700,0.74100}%
\definecolor{mycolor2}{rgb}{0.85000,0.32500,0.09800}%
\definecolor{mycolor3}{rgb}{0.92900,0.69400,0.12500}%
\begin{tikzpicture}

\begin{axis}[%
width=4.521in,
height=3.566in,
at={(0.758in,0.481in)},
scale only axis,
xmin=-20,
xmax=40,
xlabel style={font=\color{white!15!black}},
xlabel={$P$ [dB]},
ymin=0,
ymax=3.5,
ylabel style={font=\color{white!15!black}},
ylabel={Average rate [bits per channel use]},
axis background/.style={fill=white},
xmajorgrids,
ymajorgrids,
legend style={legend cell align=left, align=left, draw=white!15!black, legend pos=south east}
]

\addplot [draw=none,only marks,color=mycolor1, line width=2.0pt, mark size=4.0pt, mark=o, mark options={solid, mycolor1}]
table[row sep=crcr]{%
	-20	0.0747875869274139\\
};
\addlegendentry{Joint Encoding (Capacity)}

\addplot [draw=none,only marks,color=mycolor3, line width=2.0pt, mark size=4.0pt, mark=asterisk, mark options={solid, mycolor3}]
table[row sep=crcr]{%
	-20	0.0515603497624397\\
};
\addlegendentry{Layered Encoding}

\addplot [draw=none,only marks,color=mycolor2, line width=2.0pt, mark size=2.8pt, mark=square, mark options={solid, mycolor2}]
table[row sep=crcr]{%
	-20	0.0315525159239769\\
};
\addlegendentry{Max-SNR}

\addplot [color=mycolor1, line width=2.0pt, mark size=4.0pt, mark=o, mark options={solid, mycolor1}]
  table[row sep=crcr]{%
-20	0.0747875869274139\\
-15	0.215550720691681\\
-10	0.52032208442688\\
-5	1.04024302959442\\
0	1.72782444953918\\
5	2.46476340293884\\
10	3.04373049736023\\
15	3.36393857002258\\
20	3.43495750427246\\
25	3.46469807624817\\
30	3.46328163146973\\
35	3.46755981445313\\
40	3.46729707717896\\
};

\addplot [color=mycolor2, line width=2.0pt, mark size=2.8pt, mark=square, mark options={solid, mycolor2}]
  table[row sep=crcr]{%
-20	0.0315525159239769\\
-15	0.0979757383465767\\
-10	0.258227527141571\\
-5	0.57000321149826\\
0	1.02749562263489\\
5	1.54624879360199\\
10	1.89008271694183\\
15	1.98498594760895\\
20	1.99970281124115\\
25	2\\
30	2\\
35	2\\
40	2\\
};

\addplot [color=mycolor3, line width=2.0pt, mark size=4.0pt, mark=asterisk, mark options={solid, mycolor3}]
  table[row sep=crcr]{%
-20	0.0515603497624397\\
-15	0.146158263087273\\
-10	0.367109328508377\\
-5	0.773833870887756\\
0	1.33042454719543\\
5	1.87459242343903\\
10	2.26059079170227\\
15	2.43572354316711\\
20	2.48659062385559\\
25	2.4965283870697\\
30	2.49940371513367\\
35	2.49999928474426\\
40	2.5\\
};

\addplot [color=mycolor1, dashed, line width=2.0pt, mark size=4.0pt, mark=o, mark options={solid, mycolor1}]
  table[row sep=crcr]{%
-20	0.11994735147804\\
-15	0.337321032322943\\
-10	0.80458310611546\\
-5	1.5048186403513\\
0	2.24544056236744\\
5	2.7397147500515\\
10	2.93404737949371\\
15	2.98443801403046\\
20	2.99592007398605\\
25	2.99865576267242\\
30	2.99988775730133\\
35	3.00000027894974\\
40	3.00000028133392\\
};

\addplot [color=mycolor2, dashed, line width=2.0pt, mark size=2.8pt, mark=square, mark options={solid, mycolor2}]
  table[row sep=crcr]{%
-20	0.120283286515623\\
-15	0.337958486527205\\
-10	0.804675507247448\\
-5	1.44884290516377\\
0	1.88779371321201\\
5	1.99151101589203\\
10	1.99977520346642\\
15	2\\
20	2\\
25	2\\
30	2\\
35	2\\
40	2\\
};

\addplot [color=mycolor3, dashed, line width=2.0pt, mark size=4.0pt, mark=asterisk, mark options={solid, mycolor3}]
  table[row sep=crcr]{%
-20	0.0913286153320223\\
-15	0.244475718922913\\
-10	0.573546159565449\\
-5	1.12352753281593\\
0	1.74290596067905\\
5	2.17941991448402\\
10	2.39279752731323\\
15	2.46967225551605\\
20	2.49252441167831\\
25	2.49735090494156\\
30	2.49946975708008\\
35	2.49999945640564\\
40	2.5\\
};

\end{axis}
\end{tikzpicture}%
	}
	\caption{Average rate vs power constant $P$ for $N=2$, $K=3$, $A=2$, $m=2$, and $\tau=1$. Solid and dashed lines are for 4-ASK and QPSK input constellations, respectively.}
	\label{fig:cap_vs_P}
\end{figure}
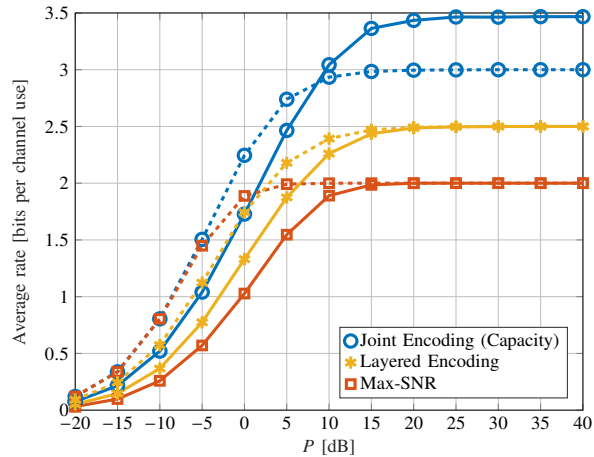

In \cref{fig:cap_vs_P} we plot the average rate as a function of the average power $P$, with $N=2$ receiver antennas, $K=3$ RIS elements, $A=2$ available phase shifts, a symbol-to-RIS control rate $m=2$, and input constellation given by QPSK $\mathcal B=\{\pm\sqrt{P},\pm i\sqrt{P}\}$ or 4-ASK $\mathcal B=\{\beta,3\beta,5\beta,7\beta\}$ with $\beta=\sqrt{P/21}$.
For layered coding, we set $\tau=1$ pilot, which was seem to maximize the rate in this experiment.
For very low SNR, i.e., less than $-20$dB, it is observed that the max-SNR approach is close to being optimal, and hence, in this regime, encoding information in the RIS configuration does not increase the rate. For larger SNR levels of practical interest, however, joint encoding provides significant gain over the max-SNR scheme and the layered approach proposed in \cref{sec:layered}, with the latter in turn strictly improving over the max-SNR scheme. In this regard, we note that we have numerically verified that, in this regime, the optimal joint distribution $p(\mathbf{x},\pmb{\theta})$ in~\eqref{eq:capacity} is not a product distribution except for very large SNR levels (see discussion after \cref{prop:high_SNR}).
Finally, while PSK outperforms ASK when used with the max-SNR and layered-encoding schemes, the opposite is true with joint encoding in the high-SNR regime. In fact, as discussed in \cref{prop:high_SNR}, in the high-SNR regime, out of all finite input sets $\mathcal B$ with the same size, ASK achieves the maximum capacity.

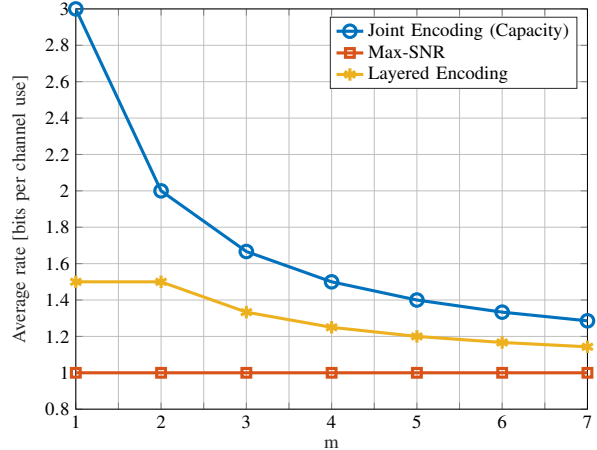
\begin{figure}[!t]
	\centering
	\resizebox {0.5\columnwidth} {!} {	
%
%
\definecolor{mycolor1}{rgb}{0.00000,0.44700,0.74100}%
\definecolor{mycolor2}{rgb}{0.85000,0.32500,0.09800}%
\definecolor{mycolor3}{rgb}{0.92900,0.69400,0.12500}%
\begin{tikzpicture}

\begin{axis}[%
width=4.521in,
height=3.566in,
at={(0.758in,0.481in)},
scale only axis,
xmin=1,
xmax=7,
xlabel style={font=\color{white!15!black}},
xlabel={m},
ymin=0.8,
ymax=3,
ylabel style={font=\color{white!15!black}},
ylabel={Average rate [bits per channel use]},
axis background/.style={fill=white},
xmajorgrids,
ymajorgrids,
xticklabels={,1,,2,,3,,4,,5,,6,,7}, 
legend style={legend cell align=left, align=left, draw=white!15!black}
]
\addplot [color=mycolor1, line width=2.0pt, mark size=4.0pt, mark=o, mark options={solid, mycolor1}]
  table[row sep=crcr]{%
1	3\\
2	2\\
3	1.66666666666667\\
4	1.5\\
5	1.4\\
6	1.33333333333333\\
7	1.28571428571429\\
};
\addlegendentry{Joint Encoding (Capacity)}

\addplot [color=mycolor2, line width=2.0pt, mark size=2.8pt, mark=square, mark options={solid, mycolor2}]
  table[row sep=crcr]{%
1	1\\
2	1\\
3	1\\
4	1\\
5	1\\
6	1\\
7	1\\
};
\addlegendentry{Max-SNR}

\addplot [color=mycolor3, line width=2.0pt, mark size=4.0pt, mark=asterisk, mark options={solid, mycolor3}]
  table[row sep=crcr]{%
1	1.5\\
2	1.5\\
3	1.33333333333333\\
4	1.25\\
5	1.2\\
6	1.16666666666667\\
7	1.14285714285714\\
};
\addlegendentry{Layered Encoding}

\end{axis}

\end{tikzpicture}%
	}
	\caption{Average rate vs the RIS control rate factor $m$ for $N=2$, $K=2$, $A=2$, $P=40$ dB, $\tau=1$, and 2-ASK input constellation.}
	\label{fig:cap_vs_DT_2PAM_A=2_P=40}
\end{figure}

The gain of using the state of the RIS as a medium for conveying information is expected to decrease as the rate of the control link from transmitter to RIS decreases. This is illustrated in \cref{fig:cap_vs_DT_2PAM_A=2_P=40}, where we plot the average rate as a function of the RIS control rate factor $m$, with $N=2$ receiver antennas, $K=2$ RIS elements, $A=2$ available phase shifts, an average power constraint of $P=40$ dB, the input constellation 2-ASK, and $\tau=1$ pilot for layered encoding. 
Note that the performance of the layered-encoding scheme is identical for $m=1$ and $m=2$ due to the use of the pilot symbol, as described in \cref{sec:layered}. It is observed that, while, for $m=1$, joint encoding achieves three times the rate of max-SNR, the gain reduces to a factor of $1.3$ for $m=7$.

\section{Conclusions}
In this work, we have studied the capacity of a Reconfigurable Intelligent Surface (RIS)-aided channel. Focusing on a fundamental model with one transmitter and one receiver, the common approach of using the RIS as a passive beamformer to maximize the SNR at the receiver was shown to be generally suboptimal in terms of the achievable rate for finite input constellations. Instead, the capacity-achieving scheme was proved to jointly encode information in the RIS configuration as well as in the transmitted signal. In addition, a suboptimal, yet practical, strategy based on layered encoding and successive cancellation decoding was demonstrated to outperform passive beamforming for sufficiently high SNR levels. 

Among related open problems, we mention the design of low-complexity joint encoding and decoding strategies that approach capacity, the derivation of the capacity for channels with imperfect a priori CSI \cite{lin2019channel,wang2019channel,you2019intelligent} or noisy RIS \cite{badiu2019communication}, and extensions to RIS systems with multiple users/surfaces \cite{huang2018energy,guo2019weighted,zhou2019intelligent,ntontin2019reconfigurable} or with security constraints \cite{guan2019intelligent,feng2019secure,lu2019intelligent}.

\appendix
\subsection{Proof of Proposition \ref{prop:capacity}}\label{app:capacity}
The model~\eqref{eq:channel} can be viewed as a standard channel with input $(\mathbf{x},\pmb{\theta})$ and output $\mathbf{Y}$. This is because the transmitter directly controls the states of the RIS $\mathbf{S}(t)$ for $t\in[n/m]$. Therefore, it follows from the channel coding theorem \cite[Ch. 7]{cover2006elements} that the capacity can be expressed as
\begin{IEEEeqnarray}{c}\label{eq:app_capacity}
	C(\mathbf{g},\mathbf{H})=\max_{\substack{
			p(\mathbf{x},\pmb{\theta}):\\
			\mathbb E[|x_i|^2]\leq P,\\
			\mathbf{x}\in\mathcal B^m,~\pmb{\theta}\in\mathcal A^K
	}}\frac{1}{m}\sqb{h(\mathbf{Y})-h(\mathbf{Y}|\mathbf{x},\pmb{\theta})}.
\end{IEEEeqnarray}
Since the conditional probability density function of the output $\mathbf{Y}$ given the input $(\mathbf{x},\pmb{\theta})$ is
\begin{IEEEeqnarray}{c}
	p(\mathbf{Y}|\mathbf{x},\pmb{\theta})=\frac{1}{\pi^{Nm}}e^{-\left\lVert\mathbf{Y}-\mathbf{H}\mathbf{S}\mathbf{g}\mathbf{x}\right\rVert^2_\text{F}},
\end{IEEEeqnarray}
we have $h(\mathbf{Y}|\mathbf{x},\pmb{\theta})=mN\log_2(\pi e)$ and the differential entropy $h(\mathbf{Y})$ can be written as (see, e.g., \cite[Eq. (3)]{ungerboeck1982Channel} and \cite[Eq. (4)]{he2005Computing})
\begin{IEEEeqnarray}{c}\label{eq:diff_entropy}
	h(\mathbf{Y})=mN\log_2(\pi)
	-\sum_{\mathbf{x}\in\mathcal B^m}\sum_{\pmb{\theta}\in\mathcal A^K}p(\mathbf{x},\pmb{\theta})\int_{\mathbb C^{N\times m}}\frac{1}{\pi^{Nm}}e^{-\lVert \mathbf{Z}\rVert^2_\text{F}}f_\text{c}(\mathbf{x},\pmb{\theta},\mathbf{Z})~\text{d}\mathbf{Z},
\end{IEEEeqnarray}
where the function $f_\text{c}(\mathbf{x},\pmb{\theta},\mathbf{Z})$ is defined in~\eqref{eq:def_f_joint}. Note that the differential entropy $h(\mathbf{Y})$ is a concave function of $p(\mathbf{x},\pmb{\theta})$ for fixed $p(\mathbf{Y}|\mathbf{x},\pmb{\theta})$ \cite[Theorem 2.7.4]{cover2006elements}, Therefore, problem~\eqref{eq:app_capacity} can be solved using standard tools.

\subsection{Proof of Proposition \ref{prop:max_snr}}\label{app:maxSNR_discrete}
Fix a phase shift vector $\pmb{\theta}$ and an input distribution $p(x)$. Since the state of the RIS is constant for the entire transmission, the channel~\eqref{eq:channel} can be restated as
\begin{IEEEeqnarray}{c}
	\mathbf{y}(t)=\mathbf{H}\mathbf{S}\mathbf{g}x(t)+\mathbf{z}(t),
\end{IEEEeqnarray}
where $\mathbf{z}(t)\sim\mathcal{CN}(0,\mathbf{I}_N)$. For this channel, by \cite[Ch. 7]{cover2006elements}, the following rate is achievable
\begin{IEEEeqnarray}{c}\label{eq:maxSnr_base}
	R(\mathbf{g},\mathbf{H})=I(x;\mathbf{y})
	=
	h(\mathbf{y})-N\log_2(\pi e),
\end{IEEEeqnarray}
where, similar to the proof of \cref{prop:capacity}, the differential entropy $h(\mathbf{y})$ can be expressed as 
\begin{IEEEeqnarray}{c}\label{eq:diffEntropy}
	h(\mathbf{y})=N\log_2(\pi)-\sum_{x\in\mathcal B}p(x)\mathbb E \sqb{f_\text{max-SNR}(x,\pmb{\theta},\mathbf{z})},\IEEEeqnarraynumspace
\end{IEEEeqnarray}
with the function $f_\text{max-SNR}(x,\pmb{\theta},\mathbf{z})$ defined in \eqref{eq:def_f(x,z)}.

\subsection{Proof of Proposition \ref{prop:high_SNR}}\label{app:high_SNR}
In the high-SNR regime, we can obtain the limit
\begin{IEEEeqnarray}{c}
	I(\mathbf{x},\pmb{\theta};\mathbf{Y})=I(\mathbf{C};\mathbf{Y})
	\xrightarrow[P\rightarrow\infty]{}
	H(\mathbf{C})
	\leq
	\log_2|\mathcal C|,
\end{IEEEeqnarray}
where equality is achieved for a uniform input distribution $p(\mathbf{x},\pmb{\theta})$.

\subsection{Proof of Proposition \ref{prop:ach_layered_discrete}}\label{app:ach_layered}
The transmitted message $w$ is divided into two layers $w_1$ and $w_2$ that are encoded in the phase shift vector $\pmb{\theta}(t)$ and the transmitted signal $\mathbf{x}(t)$, respectively, for $t\in [n/m]$.
Let $\mathbf{y}_i(t)$ denote the $i$th column of the received signal matrix $\mathbf{Y}(t)$~\eqref{eq:channel}, $i\in[m]$, which can be expressed as
\begin{IEEEeqnarray}{c}
	\mathbf{y}_i(t)=\mathbf{H}\diag\left(e^{j\theta_1(t)},\ldots,e^{j\theta_K(t)}\right)\mathbf{g}x_i(t)+\mathbf{z}_i(t),
\end{IEEEeqnarray}
where the white Gaussian noise $\mathbf{z}_i(t)\sim\mathcal{CN}(0,\mathbf{I}_N)$ denotes the additive noise at the receiving antennas during the transmission of the $i$th symbol in the $t$th block.
By fixing pilots $x_i(t)=\sqrt{P}$ for all $i\in[\tau]$ and $t\in[n/m]$, we obtain
\begin{IEEEeqnarray}{rCl}\label{eq:psk_ch}
	\frac{1}{\tau}\sum_{i=1}^{\tau}\mathbf{y}_i(t)&=&\mathbf{H}\diag\left(e^{j\theta_1(t)},\ldots,e^{j\theta_K(t)}\right)\mathbf{g}\sqrt{P}+\frac{1}{\tau}\sum_{i=1}^{\tau}\mathbf{z}_i(t)\IEEEnonumber\\
	&=&
	\mathbf{H}\diag(\mathbf{g})\tilde{\mathbf{x}}(t)+\tilde{\mathbf{z}}(t),
\end{IEEEeqnarray}
where we have defined vectors $\tilde{\mathbf{z}}(t)\triangleq \sum_{i=1}^{\tau}\mathbf{z}_i(t)/\tau\sim\mathcal{CN}(0,\mathbf{I}_N/\tau)$ and $\tilde{\mathbf{x}}(t)\triangleq(\sqrt{P}e^{j\theta_1(t)},\ldots,\sqrt{P}e^{j\theta_K(t)})^\intercal$. The channel~\eqref{eq:psk_ch} is equivalent to a point-to-point Gaussian MIMO channel with PSK input and an average power constraint of $P$ in which a precoder given by $\diag(\mathbf{g})$ is applied, and the noise at each receiving antenna has variance $1/\tau$. Therefore, it follows from \cite[Eq. (4)]{he2005Computing} that a rate of $R_1(\mathbf{g},\mathbf{H},\tau)/\tilde{m}$ is achievable for layer $w_1$. The factor $\tilde{m}=\max\{\tau+1,m\}$ is due to the fact that this scheme requires a transmitted block of size greater than $\tau$ in order to encode $w_2$ in the symbols of $\mathbf{x}(t)$ excluding the first $\tau$ symbols.

Once the receiver decodes the first layer $w_1$, it knows the state of the RIS $\mathbf{S}(t)$ for all $t\in[n/m]$. Therefore, the $ith$ column of the received signal matrix, for $i\geq \tau+1$, can be written as
\begin{IEEEeqnarray}{c}\label{eq:fast_fade_ch}
	\mathbf{y}_i(t)=\mathbf{H}\mathbf{S}(t)\mathbf{g}x_i(t)=\tilde{\mathbf{g}}(t)x_i(t),\quad i=\tau+1,\ldots,\tilde{m},
\end{IEEEeqnarray}
with time-varying channel vector $\mathbf{g}(t)\triangleq \mathbf{H}\mathbf{S}(t)\mathbf{g}$, which is known to both transmitter and receiver. The channel~\eqref{eq:fast_fade_ch} is equivalent to a fast-fading model with CSI at the transmitter and receiver \cite{goldsmith1997Capacity}. Therefore, the following scheme is applied. 
The transmitter uses a uniform input distribution, i.e., $p(x)=1/|\mathcal B|$ for all $x\in\mathcal B$. 
For an RIS configuration $\pmb{\theta}$, the transmitter amplifies the transmitted symbol by $\alpha(\pmb{\theta})$ in~\eqref{eq:power_allocation}. Note that the average-power constraint is satisfied due to equality~\eqref{eq:avg_power_eq}. Hence, a rate of $I(x;\mathbf{y}|\pmb{\theta})/m$ can be achieved, where, similar to the proofs of \cref{prop:capacity} and \cref{prop:max_snr}, the mutual information $I(x;\mathbf{y}|\pmb{\theta})$ can be expressed as $R_2(\mathbf{g},\mathbf{H})$ in~\eqref{eq:tildeR2}. Since $\tilde{m}-\tau$ of the $\tilde{m}$ symbols in block $\mathbf{x}(t)$ are used for conveying information, the total rate achieved in the second layer is $(\tilde{m}-\tau)R_2(\mathbf{g},\mathbf{H})/\tilde{m}$. Thus, using both layers, the rate $R_\text{layered}(\mathbf{g},\mathbf{H},\tau)$ in~\eqref{eq:ach_layered_discrete} is achievable.

\nocite{renzo2019Reflection}

\bibliographystyle{IEEEtran}
\bibliography{IEEEabrv,myBib}

\end{document}